\documentclass{iaus}
\usepackage{graphicx}
\usepackage{natbib}
\usepackage{bm}
\graphicspath{{./fig/}{./png/}}

\def\Rm{\mbox{\rm Re}_{\rm M}}
    
\title[Decay of trefoil and other magnetic knots]
{Decay of trefoil and other magnetic knots}

\author[S. Candelaresi, F. Del Sordo, \& A. Brandenburg]
{Simon Candelaresi,
 Fabio Del Sordo
 \and Axel Brandenburg}

\affiliation{NORDITA, AlbaNova University Center,
Roslagstullsbacken 23, SE-10691 Stockholm, Sweden and \\
Department of Astronomy, Stockholm University,
SE 10691 Stockholm, Sweden}

\pubyear{2010}
\volume{274}  
\pagerange{461--463}
\jname{Advances in Plasma Astrophysics}
\editors{A. Bonanno, E. de Gouveia dal Pino \& A. Kosovichev, eds.}
\doi{10.1017/S1743921311007496}
\begin{document}

\maketitle

\begin{abstract}
Two setups with interlocked magnetic flux tubes are
used to study the evolution of magnetic energy and
helicity on magnetohydrodynamical (MHD) systems
like plasmas.
In one setup the initial helicity is zero while in the other it is finite.
To see if it is the actual linking or merely the helicity content that
influences the dynamics of the system we also consider a setup with unlinked field lines
as well as a field configuration in the shape of a trefoil knot.
For helical systems the decay of magnetic energy is slowed down by
the helicity which decays slowly.
It turns out that it is the helicity content, rather than the actual linking,
that is significant for the dynamics.
\end{abstract}

Magnetic helicity has been shown to play an important role in the dynamo
process \citep{BBSubReview2005}. For periodic systems where helicity is conserved
simulations have shown that with increasing
magnetic Reynolds number $\Rm$ the saturation magnetic field strength
decreases like $\Rm^{-1/2}$ \citep{BBDboler2001}.
This is problematic for astrophysical bodies
since for the Sun $\Rm = 10^{9}$ and galaxies $\Rm = 10^{14}$.
In order to alleviate this quenching the
magnetic helicity of the small scale fields
needs to be shed \citep{ssHelLoss09}.

In the active regions of the Sun twisted magnetic field lines have been observed \citep{Pev95}.
Later it was shown \citep{Lek96} that the magnetic field in sunspots
gets twisted before it emerges out of the surface. 
\citep{Manoharan1996} and \citep{Canfield1999} demonstrated that
helical structures 
are more likely to erupt into coronal mass ejections.
This suggests that the Sun sheds helicity.

The magnetic helicity is related to the mutual linking for two
non-intersecting flux tubes via \citep{Mof69}
$$
H = \int_{V}\bm{A}\cdot\bm{B} \ {\rm d} V = 2n\phi_{1}\phi_{2},
$$
where $H$ is the magnetic helicity, $\bm{B}=\nabla\times\bm{A}$ is the
magnetic field in terms of the vector potential $\bm{A}$,
$\phi_{1}$ and $\phi_{2}$ are the magnetic fluxes through the tubes and $n$ is the 
linking number. The flux tubes may not have internal twist.
In the limit of large $\Rm$
$H$ is a conserved quantity as well as the linking number.

In presence of magnetic helicity the magnetic energy decay is constrained
via the realizability condition \citep{Mof69} which gives a lower bound
for the spectral magnetic energy
$$
M(k) \ge k|H(k)|/2\mu_{0} \quad {\rm with} \quad
\int M(k) \ {\rm d} k = \langle\bm{B}^{2}\rangle/2\mu_{0}, \quad
\int H(k) \ {\rm d} k = \langle\bm{A}\cdot\bm{B}\rangle,
\label{eq: realizability condition}
$$
the magnetic permeability $\mu_{0}$, where $\langle.\rangle$
denotes volume integrals.

In this work we extend earlier work \citep{DSo10} where the dynamics of interlocked flux rings,
with and without helicity, was studied
as well as a non-interlocked configuration.
Here we also study
a self-interlocked flux tube in the form of a trefoil knot.

\begin{figure}[t!]
\begin{minipage}[b]{0.5\linewidth}
\centering
\includegraphics[width=0.2\linewidth]{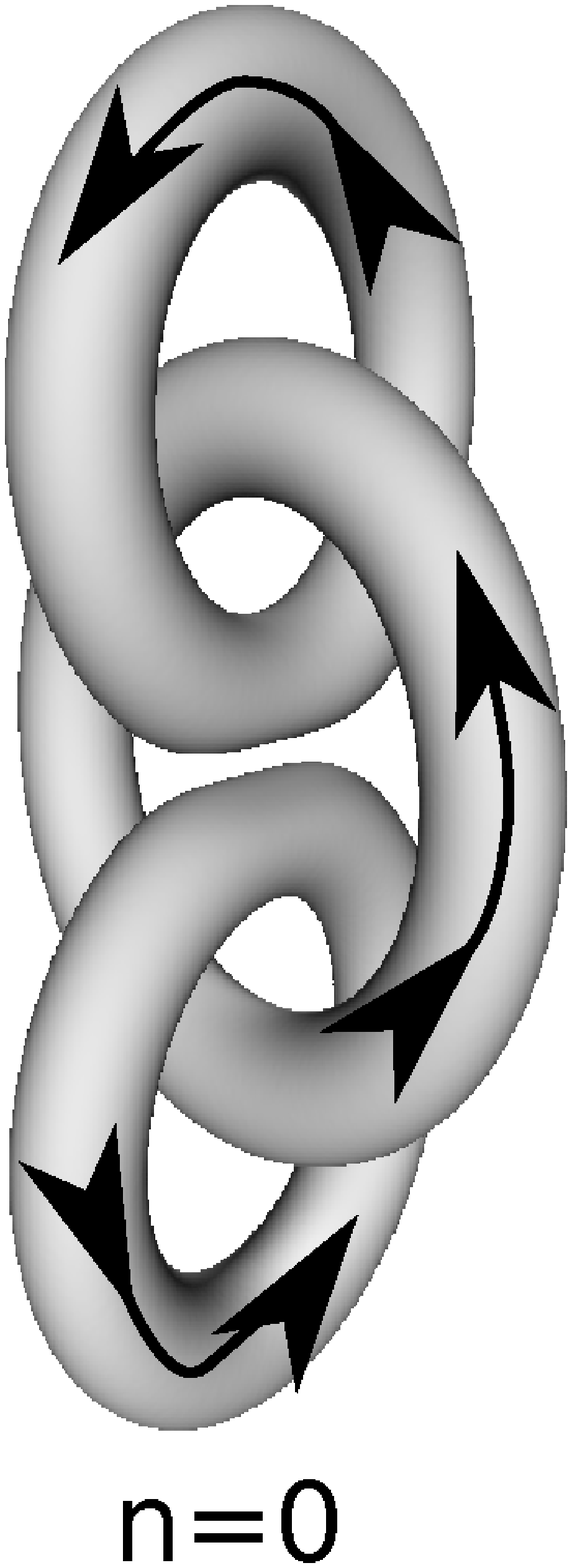} \quad
\includegraphics[width=0.2\linewidth]{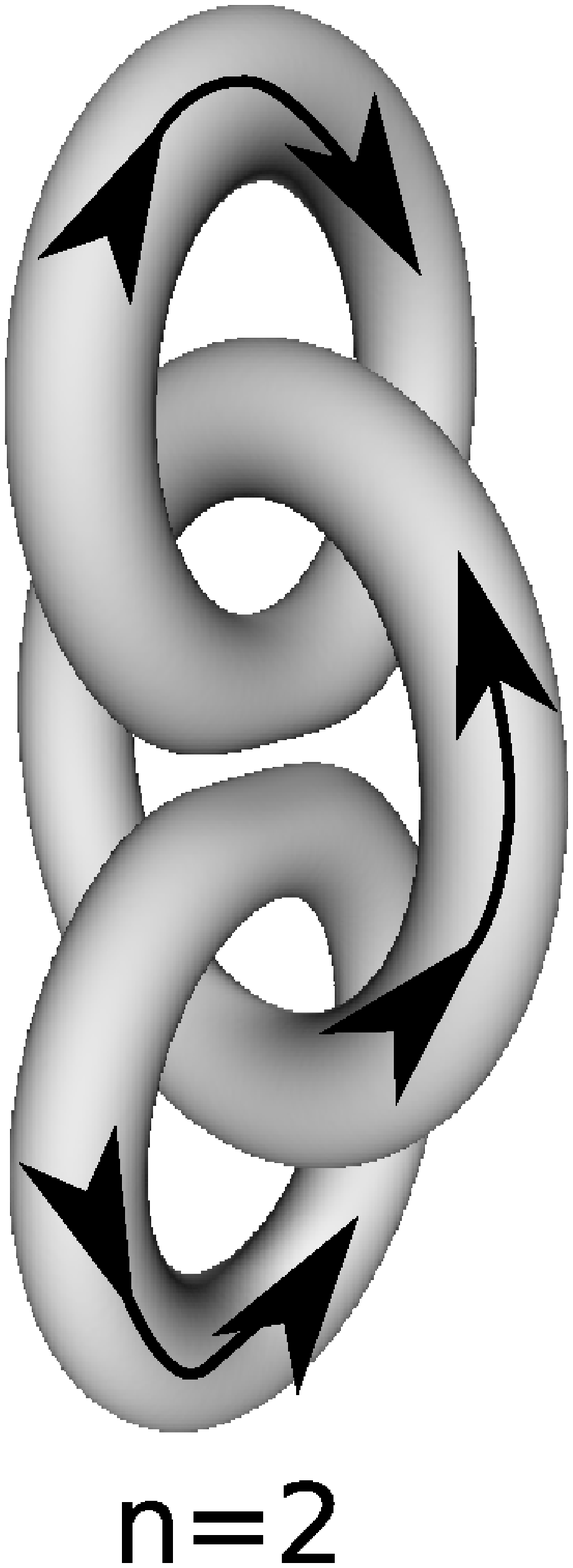} \quad
\includegraphics[width=0.3\linewidth]{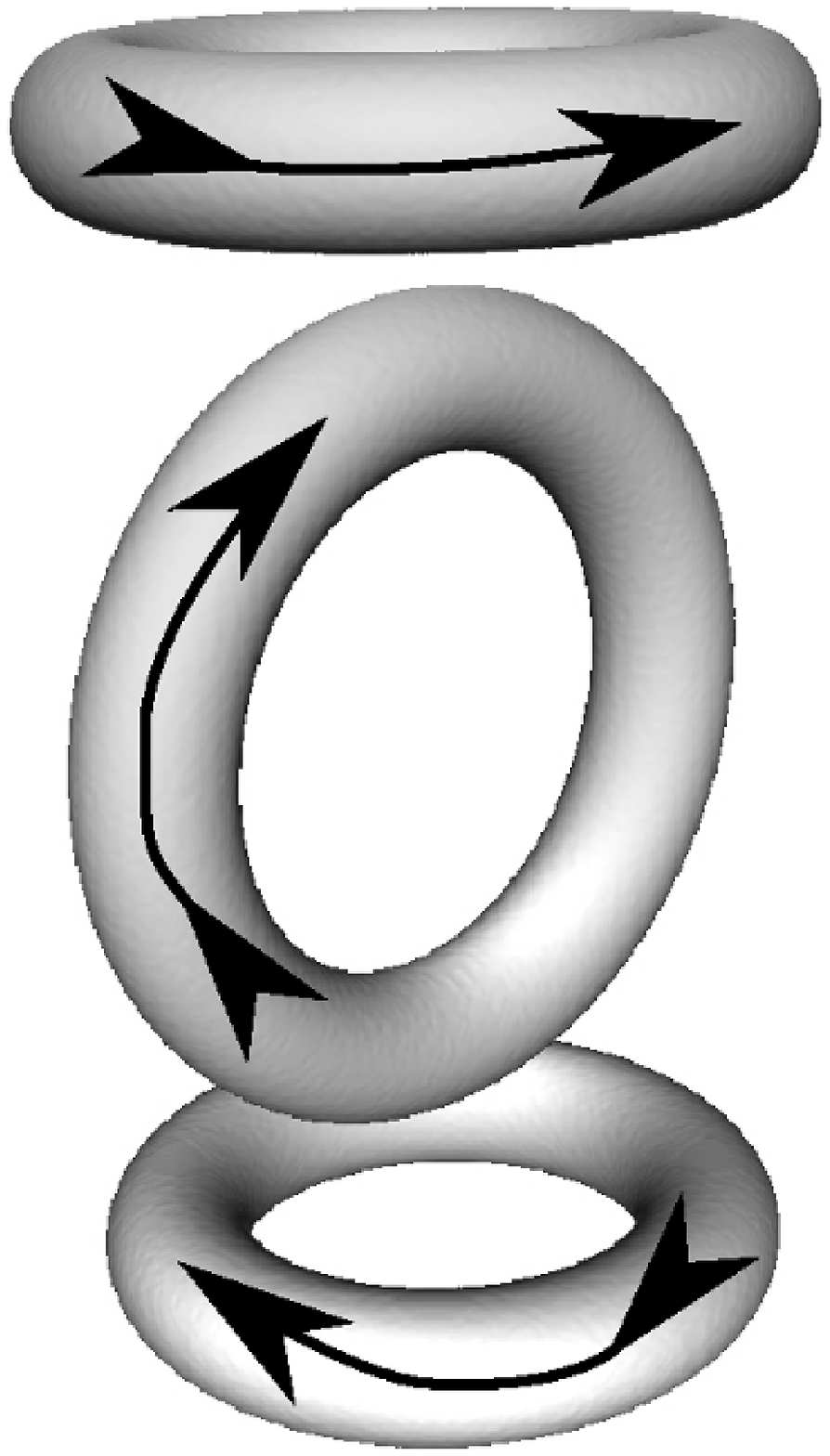}
\caption{The three triple ring configurations for the initial time.
From left to right: interlocked rings with no helicity,
interlocked rings with finite helicity and
non-interlocked rings without helicity.
The arrows indicate the direction of the magnetic field.
Adapted from \citep{DSo10}.
}\label{fig: initial configuration}
\end{minipage}
\hspace{0.5cm}
\begin{minipage}[b]{0.5\linewidth}
\centering
\includegraphics[width=0.84\linewidth]{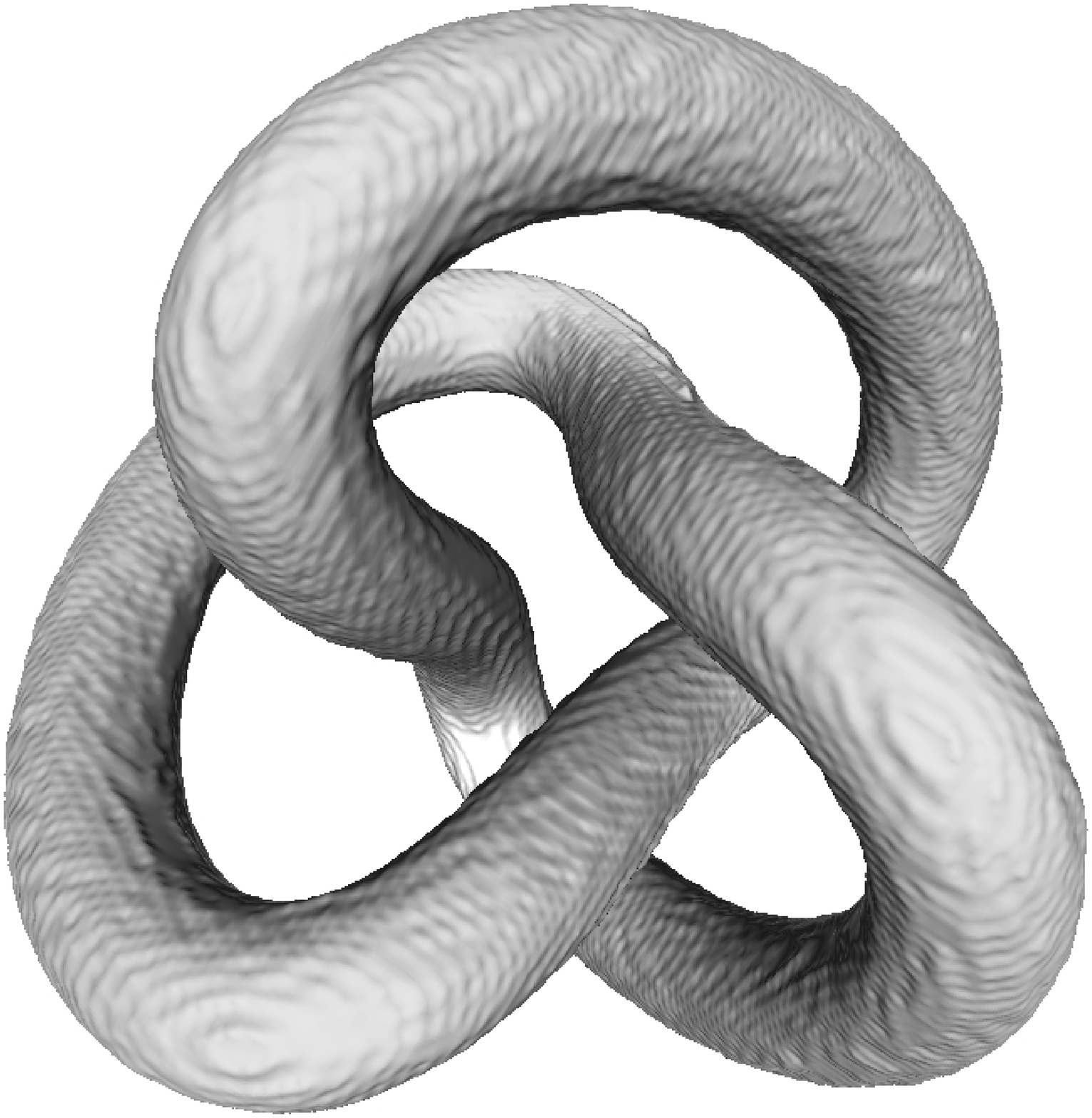}
\caption{The initial magnetic field configuration for the trefoil knot.
}\label{fig: trefoil knot}
\end{minipage}
\end{figure}

The three-rings setups consist of three magnetic flux tubes.
In two configurations they are interlocked
where in one the helicity
is zero and in the other one it has a finite value,
as shown in Fig.\,\ref{fig: initial configuration}.
In the third setup we instead consider unlocked rings. 
Since the rings do not have internal twist,
the helicity of this last configuration is zero. 
We also study the evolution of a self-interlocked flux tube having the form of a trefoil
knot, with finite helicity (Fig.\,\ref{fig: trefoil knot}).
In this case we have $H=3\phi^2$, so the linking number is $n=3/2$.
All of these setups evolve according to the full resistive equations
of MHD for an isothermal compressible medium.
The Alfv\'en time  is used as time unit.

\begin{figure}[b!]
\centering
\includegraphics[width=0.49\linewidth]{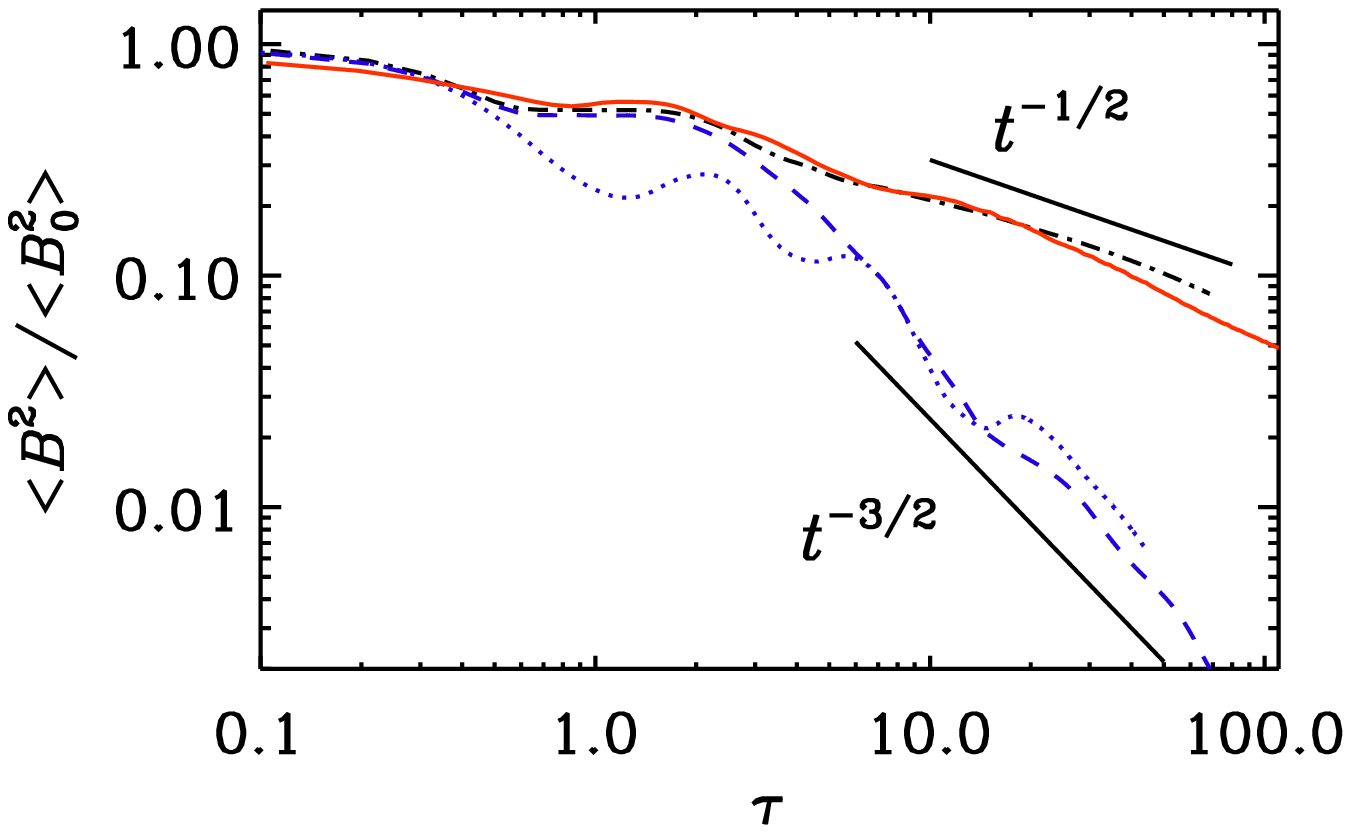}
\includegraphics[width=0.47\linewidth]{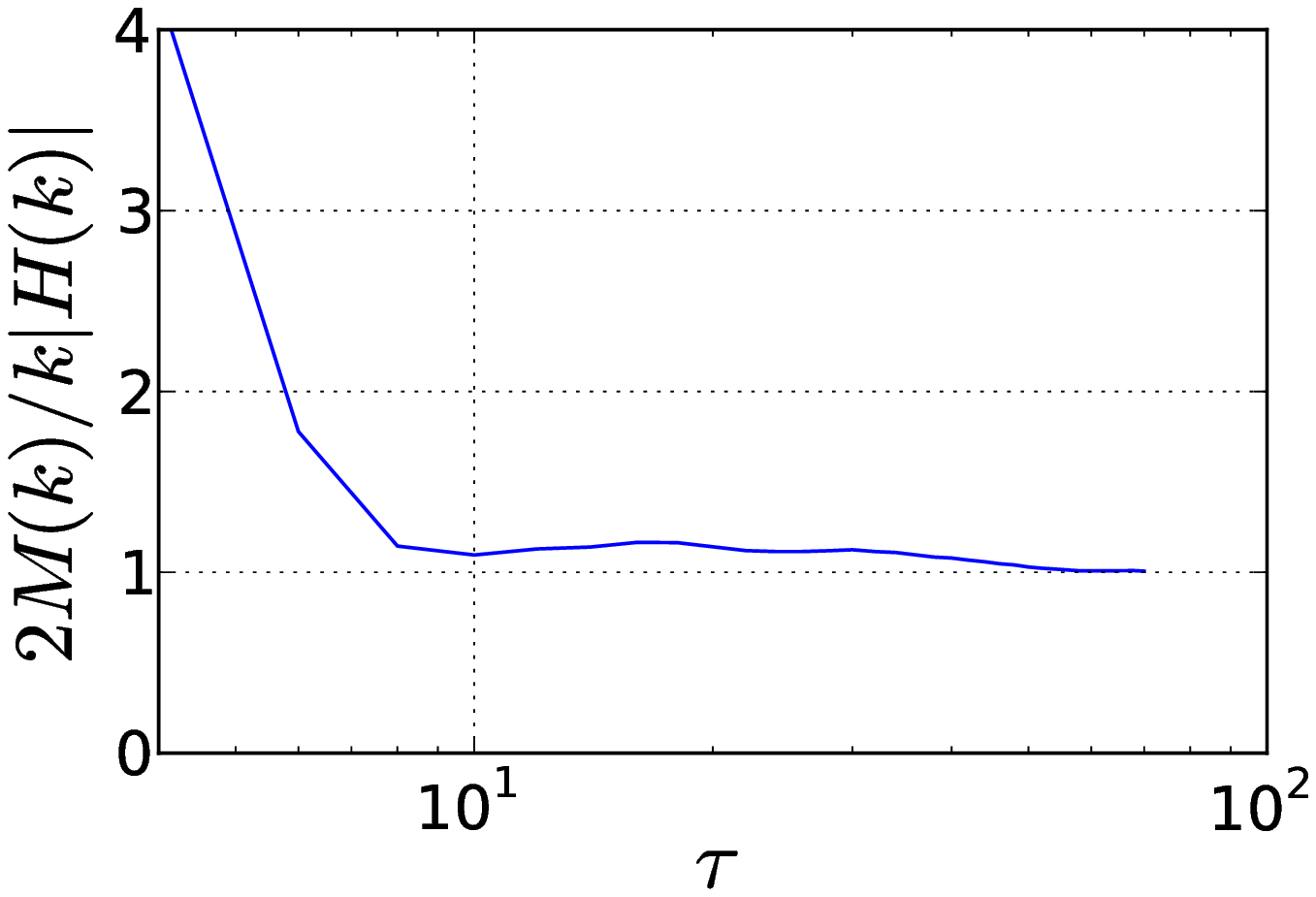}
\caption{Left: Evolution of the normalized magnetic energy for the
trefoil knot (solid/red line) compared with various
three-ring configurations with
$n=2$ (dash-dotted line), $n=0$ (dashed/blue line), and the non-interlocked
case (dotted/blue line).
Right: $\mathcal{H}(k)$ with $k=4$ for the helical triple ring configuration.
}
\label{fig: energy decay}
\end{figure}

\begin{figure}[t!]
\begin{center}
 \includegraphics[width=0.6\linewidth]{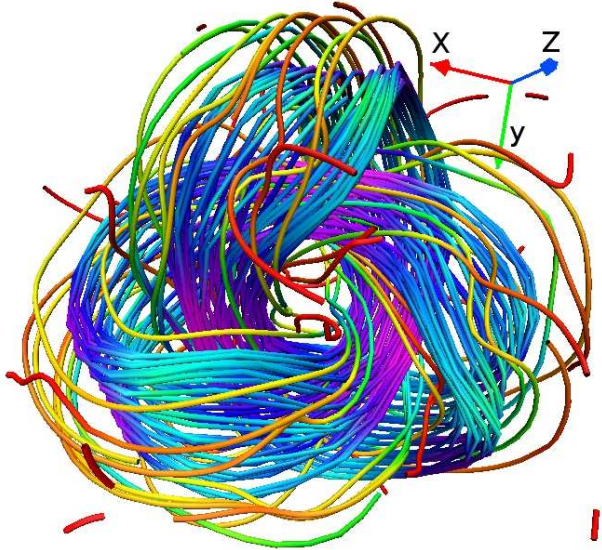}
 \caption{Magnetic field lines at 5 Alfv\'en times for the 
          trefoil knot.
          The colors represent the magnitude of the magnetic field.
Note that internal twist generation is weak.
}\label{fig: field lines evolution}
\end{center}
\end{figure}

As a consequence of the realizability condition the magnetic energy
cannot decay faster than the helicity.
To check the importance of this restriction we plot the ratio
\begin{equation}
\mathcal{H}(k) = 2\mu_{0}M(k)/k|H(k)|
\end{equation}
in Fig.\,\ref{fig: energy decay} for $k=4$, which is the typical
scale of the system. The magnetic energy at that scale reaches
its allowed minimum value already at $\tau = 10$, which confirms
the importance of the realizability condition in this decay process.
The setups with finite $H$ show a slower decay than the setups with no helicity
(Fig.\,\ref{fig: energy decay}).
The decay of the trefoil knot follows approximately the same decay law
as the other configuration consisting of three rings with finite $H$.
Within the simulation time $H$ decays only to about one half
of the initial value
conserving then the topology.
During later times field lines reconnect and the helicity seems to go
into internal twist, which is topologically equivalent to linking;
see Fig.~\ref{fig: field lines evolution}.

The slow decay of $H$ conserves the topology of the system.
The linking is then eventually transformed into internal twisting
during magnetic reconnection.
Since both non-helical setups evolve similarly we conclude that it is
mainly the magnetic helicity and not the actual linking which influences the
dynamics.
The helical trefoil knot evolves in a similar manner.
This confirms the hypothesis that the decay of interlinked
flux structures is governed by magnetic helicity and that
higher-order invariants, advocated for example by \citep{YHW10},
may not be essential for describing this process.

In conclusion, we can say that magnetic helicity is decisive in
controlling the decay of interlocked magnetic flux structures.
If the magnetic helicity is zero, resistive decay will be fast
while with finite magnetic helicity the decay will be slow and
the speed of decay of magnetic energy depends on the speed at which
magnetic helicity decays.
This is likely an important aspect also in magnetic reconnection
problems that has not yet received sufficient attention.

\end{document}